\documentclass{article}
\usepackage{spconf,amsmath,graphicx}

\title{FB-MSTCN: A Full-Band Single-Channel Speech Enhancement Method Based on Multi-Scale Temporal Convolutional Network}
\name{Zehua Zhang$^{*}$, Lu Zhang$^{*}$, \thanks{${*}$ Equal contribution} Xuyi Zhuang, Yukun Qian, Heng Li, Mingjiang Wang}
\address{Harbin Institute of Technology, Shenzhen, China}
\begin{document}
%
\maketitle
\begin{abstract}
In recent years, deep learning-based approaches have significantly improved the performance of single-channel speech enhancement. However, due to the limitation of training data and computational complexity, real-time enhancement of full-band (48 kHz) speech signals is still very challenging. Because of the low energy of spectral information in the high-frequency part, it is more difficult to directly model and enhance the full-band spectrum using neural networks. To solve this problem, this paper proposes a two-stage real-time speech enhancement model with extraction-interpolation mechanism for a full-band signal. The 48 kHz full-band time-domain signal is divided into three sub-channels by extracting, and a two-stage processing scheme of `masking + compensation' is proposed to enhance the signal in the complex domain. After the two-stage enhancement, the enhanced full-band speech signal is restored by interval interpolation. In the subjective listening and word accuracy test, our proposed model achieves superior performance and outperforms the baseline model overall by 0.59 MOS and 4.0$\%$ WAcc for the non-personalized speech denoising task.

\end{abstract}
\section{Introduction}
\label{sec:intro}

With the significant increase in hardware computing power in recent years, the demand for high-quality speech in real-time communication applications like video conferencing and live broadcast is also increasing. Speech enhancement techniques are essential for removing noise interference to improve speech quality. However, most of the previous studies on speech enhancement are for narrow-band (8 kHz) or wide-band (16 kHz) audio, and there are few methods for 48 kHz full-band audio. Deep learning-based speech enhancement methods \cite{R1,R2,R3} have achieved impressive performance on wide-band audio, but the lack of sufficient training data has become a major limitation for full-band deep learning speech enhancement methods. The recent 4th Microsoft Deep Noise Suppression (DNS-4) Challenge\footnote{https://www.microsoft.com/en-us/research/academic-program/deep-noise-suppression-challenge-icassp-2022/} extends efforts to full-band single-channel speech enhancement tasks with a massive training dataset and real-scenario test set.

Another difficulty of full-band speech enhancement is that the increased sampling resolution leads to the need to model larger dimensional features in the neural network model, which results in higher modeling complexity. For example, in wide-band real-time scenarios, most methods \cite{R4,R5,R6,R7} have the input frame size set to 20 ms or 32 ms, corresponding to 320 and 512 points of Fourier transform analysis, respectively. If the audio is sampled at 48 kHz, the same frame size corresponds to three times the number of Fourier transform points, which triples the dimensionality of the input features and correspondingly increases the modeling dimensionality of the neural network model. Although end-to-end modeling schemes in the time domain, such as Conv-TasNet \cite{R8} and DPRNN \cite{R9}, can not be affected by the input feature dimension, it will increase the number of processing frames, resulting in a threefold increase in the number of operations to process audio per second. Therefore, this is also a challenge for the computing power of the chip in real-time applications.

In the full-band real-time scenarios, modeling sub-band spectral features, such as RNNoise \cite{R10} and PercepNet \cite{R11}, is an effective strategy to balance the computational complexity and speech denoising performance. Bark and triangular filter banks are respectively used to compress the frequency spectrum, which retains the frequency-domain information that is more important to human perception, effectively reducing the dimension of input features and thus reducing the complexity of the neural network model. However, such methods inevitably lose some spectral details, resulting in sub-optimal performance. The recent S-DCCRN model \cite{R12} proposes a two-stage modeling scheme, which can optimize the low-band and high-band separately, and further employs a full-band processing module to smooth the output of both sub-bands. This method benefits from modeling local and global frequency information, allowing it to obtain better performance.

In this paper, we also propose a two-stage processing model to enhance the full-band signal. In the first stage, the complex ratio masks (CRMs) \cite{R1} are estimated to initially enhance the noisy signal, and then the second-stage module can further compensate the enhanced complex signal. To ensure a better temporal modeling ability of the model, we introduce two long-term modeling units with fixed and dynamic receptive fields based on the multi-scale temporal convolutional network (TCN) model \cite{R13}. In addition, we propose a novel full-band signal processing mechanism of extraction and interpolation to reduce the modeling difficulty of the full-band spectrum. Specifically, the 48 kHz time-domain noisy signal is divided into three sub-channel signals (16 kHz) in the way of interval sampling. After the two-stage enhancement, the signals of three sub-channels are restored to the 48 kHz time-domain enhanced signal by interval interpolation.

\section{FB-MSTCN Model}
\label{sec:FB-MSTCN Model}

In this paper, $\left(N_{r}, N_{i}\right)$ and $\left(S_{r}, S_{i}\right)$ are the noisy real and imaginary (RI) spectrum and the clean RI spectrum respectively. The overall diagram of proposed FB-MSTCN is shown as Fig.1. We divide the single-channel full-band signal into three sub-channel wide-band signals by extraction, where $j$ represents the index of the channel. In order to make the model have a better subjective feeling and make the model easier to converge, we adopt the complex compressed feature \cite{R14} as the input of the dynamic long-term embedding unit, which is written as $\left(N_{r}^{c}, N_{i}^{c}\right)$. \begin{figure}[hb]
\centering
	\includegraphics[scale=0.55]{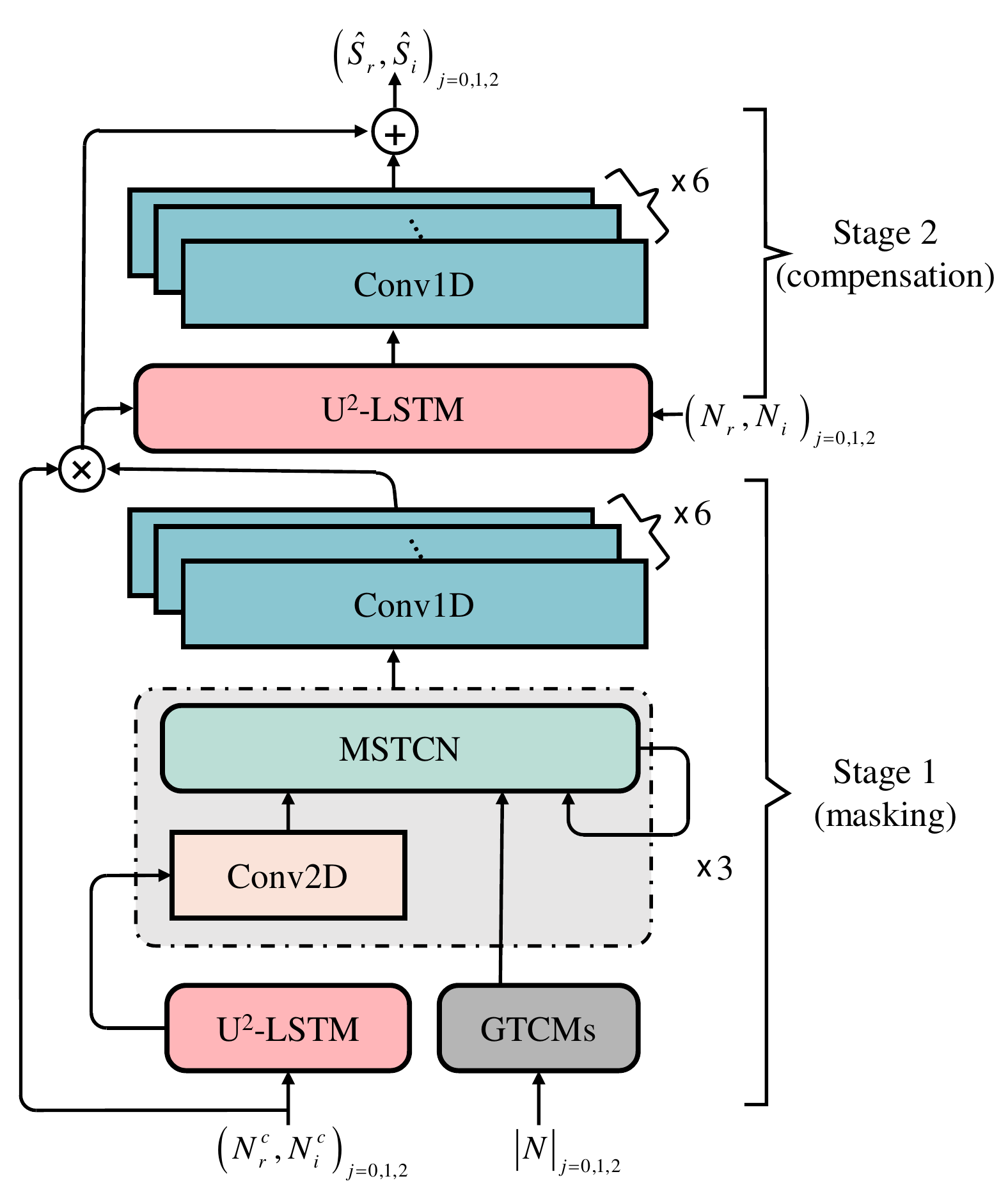}
\caption{Overall diagram of the proposed FB-MSTCN.}
\label{fig:overall}
\end{figure}
In the first stage, we use a fixed-length long-term embedding unit and a dynamic long-term embedding unit to capture the temporal dependence of speech signals, and then perform multi-scale feature analysis on them using multi-scale temporal convolution network (MSTCN) \cite{R13}. CRMs are calculated by six 1-D convolutions after MSTCN. In the second stage, a topology similar to dynamic long-term embedding unit is used to further suppress the residual noises and compensate some under-estimated spectral details. Finally, the enhanced sub-channel signals are interpolated in the time domain to obtain the final waveform. 

\subsection{Extraction and Interpolation}
\label{ssec: Extraction and Interpolation}
A novel method of extraction and interpolation in the time domain, which can be expressed as Eq.\ref{Interpolation}, is proposed to process the full-band speech signal. $n_{FB}$ represents full-band time-domain speech signal and $n_j$ represents sub-channel speech signal after extraction, where $j=0,1,2$. We expand the single channel to three sub-channels through the extraction operation, and the relationship between different sub-channels can be learned through the FB-MSTCN model. The three enhanced sub-channel speech signals can be interpolated into a full-band speech signal by the inverse operation of Eq.\ref{Interpolation}.

\begin{equation}
\label{Interpolation}
\centering
    n_{j}(m)=n_{FB}(3 \times m+j)
\end{equation}

\subsection{Fixed-Length Long-Term Embedding Unit}
\label{ssec: Fixed-Length Long-Term Embedding Unit}

In fixed-length long-term embedding unit, gated temporal convolution modules (GTCMs) are used to capture the temporal dependency information of magnitude spectrum. GTCMs as shown in Fig.2 contains four 1-D convolutions, where $k,d,c$ represent kernel size, dilation rate and output channels respectively. Each group having six GTCMs is repeated three times. In each group, the dilation rate $d$ is equal to 1, 2, 4, 8, 16, respectively, which makes the model have a fixed-length of receptive field to capture the long-term embedding feature in the magnitude domain.

\begin{figure}[htb]
\centering
	\includegraphics[scale=0.6]{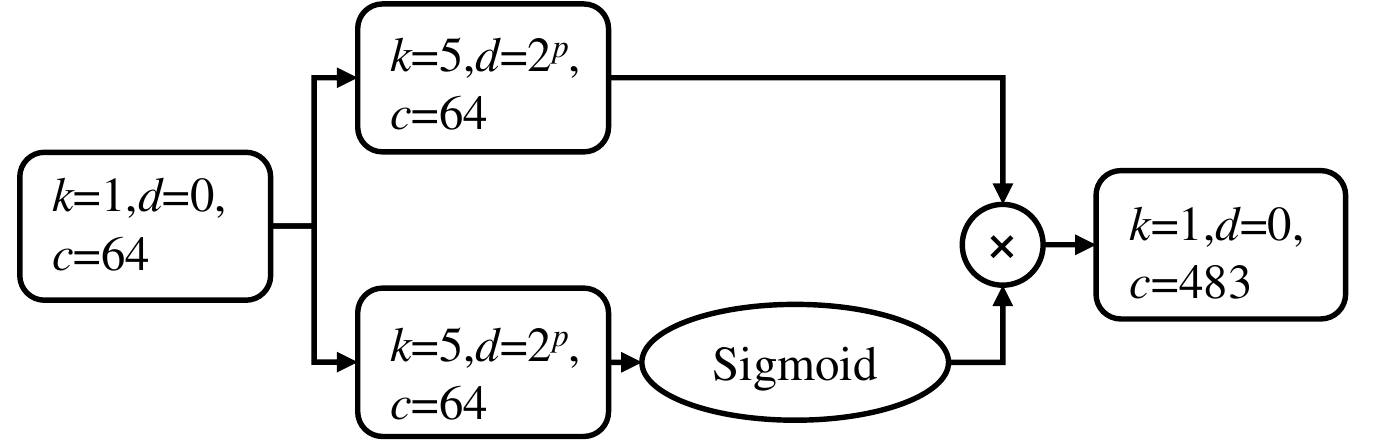}
\caption{Structure diagram of GTCMs.}
\label{fig:GTCN}
\end{figure}

\subsection{Dynamic Long-Term Embedding Unit}
\label{ssec: Dynamic Long-Term Embedding Unit}
Inspired by U$^2$-Net \cite{R15}, this paper proposes a similar topology named U$^2$-LSTM as shown in Fig.3 to model compressed complex features. On the basis of U$^2$-Net, a four-layer LSTM is added to capture dynamic long-term context information. GConv2D, GDeConv2D and IN represent gated 2-D convolution, 2-D deconvolution and InstanceNorm respectively. The convolution kernel size is (2, 5) for the first layer of GConv2D and the last layer of DeConv2D, and (2, 3) for all others. The number of channels is 64, and the stride is (1,2) in all layers. The output of the dynamic long-term embedding unit changes the number of channels from 64 to 8 as the input of MSTCN through a layer of 2-D convolution.

\begin{figure}[htb]
\centering
	\includegraphics[scale=0.6]{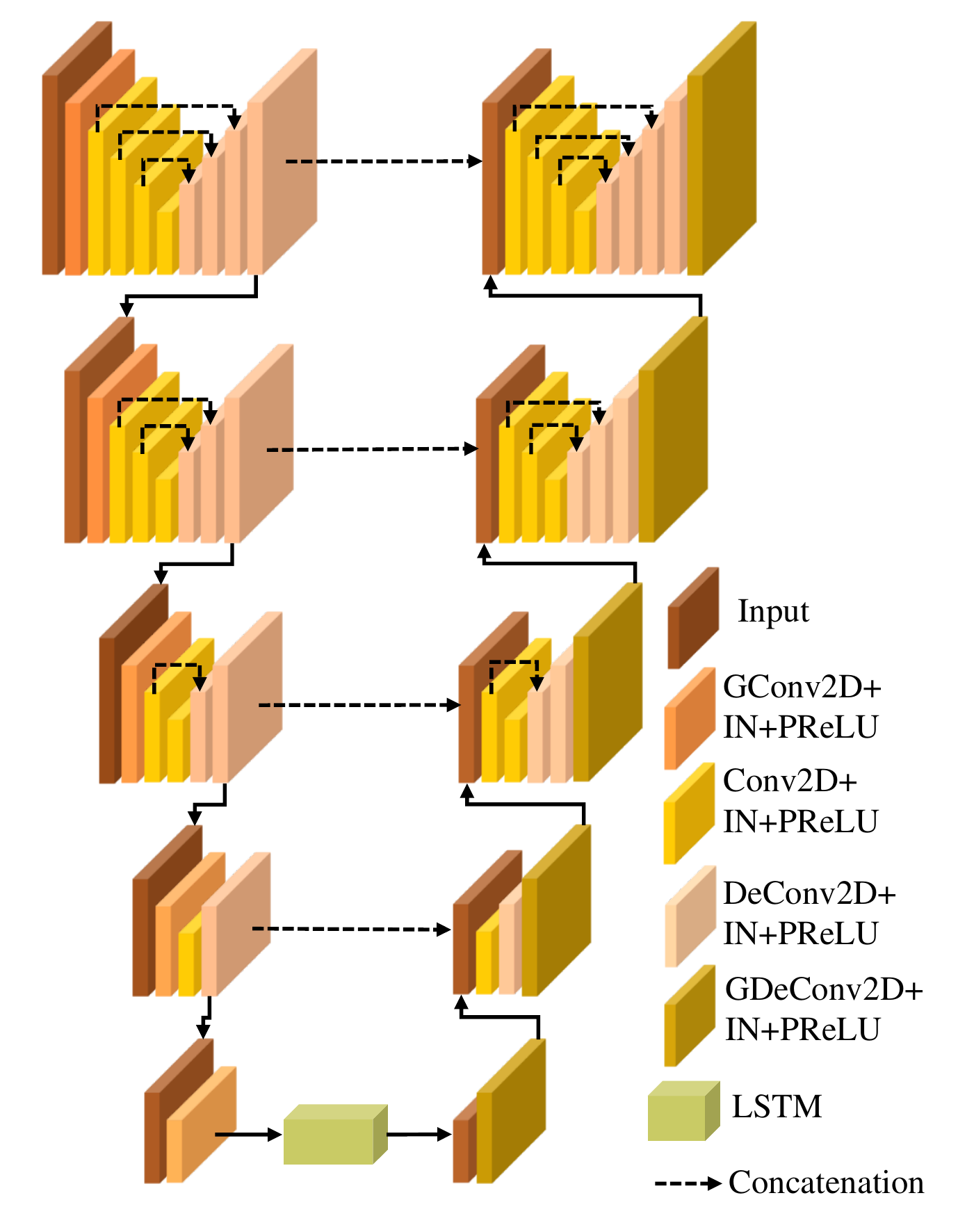}
\caption{Structure diagram of the proposed U$^2$-LSTM.}
\label{fig:U2net}
\end{figure}

\subsection{Multi-Scale TCN}
\label{ssec: multi-scale TCN}

Our previous study \cite{R13} found that increasing the granularity of the time-frequency (T-F) analysis of features contributes to improving the speech denoising performance and the generalization ability of the model. We use the proposed MSTCN framework in \cite{R13} to perform multi-scale sub-band analysis on the two obtained long-term embedding features, which can be expressed as follows:
\begin{equation}
  F\!_{m\!d\!,b\!}(\!t\!)\!=\!\left(\!Y\!\!*\!f\!_{m\!d\!,b\!}\right)\!=\!\sum_{i=0}^{K\!-\!1}\!f\!_{m\!d\!,b}(\!i\!)\!\cdot\!\left\{\!\tilde{F}\!_{m\!d\!,b\!-\!1\!},Y\!_{b}\!\right\}(\!t\!-\!d\!\cdot i)
  \label{eq1}
\end{equation}
Where \(f_{md,b}\) and \(F_{md,b}(t)\) are the multi-scale kernel and the output of current sub-band, respectively. \(t\), \(b\), \(K\) and \(d\) denote the frame index, sub-band index, kernel size and dilation factor, respectively. \(Y_b\) represents the input features of each band, \(\tilde{F}_{md,b-1}\) is the output of the adjacent band corresponding to sub-band \(b\).

The MSTCN here consists of three groups of multi-scale TCN modules, each consisting of five residual blocks with a kernel size of 3 and the dilation rate cycles in increments of 1, 3, 5, 7, and 11. In addition, the input to the MSTCN is the fixed-length long-term embedding features extracted from the GTCMs, while the dynamic long-term embedding features extracted by the U$^2$-LSTM unit are transformed into three groups of 256-dimensional forward-stacked features through the 1-D convolution.

\subsection{Compensation Model}
\label{ssec: Two-Stage Modeling}

The compensation model adopts a topology similar to dynamic long-term embedding unit, except that the number of input channels becomes 12 and the number of output channels becomes 6. The 6 channels of U$^2$-LSTM output are respectively convolved through a 1-D convolution with a kernel size of 1 to calculate the compensation value. The enhanced complex spectrum can be obtained by adding the compensation value to the masked result of the first stage. Through the compensation model, the residual noise in the first-order stage will be further suppressed, and the complex spectrum can be recovered better.

\subsection{Loss Function}
\label{ssec: Loss Function}

To ensure consistency in the optimization of the RI and magnitude spectrum, we adopt the loss function form of combined mean square error (cMSE) in \cite{R2}, as follows:
\begin{equation}
Loss=\lambda \cdot\left\|\hat{S}_{c R I}-S_{c R I}\right\|^{2}+\beta \cdot\left\|\hat{S}_{c M a g}-S_{c M a g}\right\|^{2}
\end{equation}
where $\hat{S}$ and ${S}$ represent the enhanced and ideal signals, respectively. A spectral compression method is applied for both RI and magnitude to achieve better convergence:
\begin{equation}
\hat{S}_{c M a g}=\left|\hat{S}_{M a g}\right|^{c}, \quad \hat{S}_{c R I}=\hat{S}_{c M a g} \cdot \frac{\hat{S}_{R I}}{\hat{S}_{M a g}}
\end{equation}
In this paper, $\lambda$ and $\beta$ are set to 0.3 and 0.7, respectively, and the compressed factor ${c}$ is set to 0.3. The final loss is obtained by averaging the cMSE loss over 3 sub-channels. 

\section{Experimental Results}
\label{sec:Experimental Results}

\subsection{Experimental Setup}
\label{ssec:experimental-setup}

We downloaded the full-band clean speech datasets from DNS-4 Challenge and simply cleaned the datasets to remove the speech files with low SNR. We finally got 885 hours of English data, 382 hours of German data, 130 hours of Spanish data, 127 hours of French data, 99 hours of Italian data, and 18 hours of Russian data. The noise datasets were also downloaded from the ICASSP 2022 DNS-4 datasets, for a total of 181 hours of noise data. In the end, we generated a noisy-clean training set of 100 hours for ablation study, a noisy-clean training set of 2,000 hours for DNS-4 challenge, and a test set of 5 hours to evaluate the performance of the proposed network. The SNR range is -5dB to 10dB for all noisy-clean set. Furthermore, to account for reverberation effects in real environments, all clean speech is convolved with synthetic and real room impulse responses (RIRs) provided by DNS-4 before being mixed with different noise signals. The reverberation time T$_{60}$ is between 0 and 0.8 seconds.

All the utterances are sampled at 48 kHz and cut into 30 s segments to facilitate model training. The frames are analyzed by a Hamming window with 20 ms length and 10 ms overlap. For training the two-stage network, we first train the first stage with an initial learning rate of 0.001. When the first-stage learning rate drops to 0.0005, we freeze the first-stage parameters and only update the second-stage parameters with an initial learning rate of 0.001. When the learning rate of the second-stage is also reduced to 0.0005, the parameters of the two stages are updated simultaneously. We halve the learning rate when validation loss does not decrease for consecutive 3 epochs. The Adam algorithm is used to optimize the models on every mini-batch with a batch size of 1000 consecutive input frames. Perceptual evaluation of speech quality (PESQ) \cite{R16}, short-time objective intelligibility (STOI) \cite{R17}, source to distortion ratio (SDR) \cite{R18}, and DNSMOS \cite{R19} are adopted as four evaluation metrics. Since the above four evaluation indicators are only applicable to wide-band speech signals, we downsampled the enhanced speech signals before evaluating.

\subsection{Ablation Study}
\label{ssec:ablation-study}

In this section, we conduct ablation experiments to evaluate the effectiveness of different model formulations proposed in this paper. As shown in Table 1, we compare the contribution of GTCMs, U$^2$-LSTM, compensation stage to model performance, and * represents the method of using MSTCN model to directly enhance the full-band spectrum. The results prove that the proposed extraction-interpolation processing strategy can effectively improve the DNSMOS performance. Besides, U$^2$-LSTM and compensation model contribute significantly to model performance.
\begin{table}[th]
\caption{The objective results for different models}
  \label{tab:The objective results for different models}
  \centering
\resizebox{0.98\columnwidth}{!}{
\begin{tabular}{ccccc}
\hline
Methods     & PESQ  & STOI  & SDR    & DNSMOS \\ \hline
Noisy       & 1.652 & 0.858 & 10.526 & 2.973  \\
MSTCN$^*$       & 2.257 & 0.885 & 15.853 & 3.278  \\
MSTCN       & 2.267 & 0.890 & 15.227 & 3.317  \\
GTCMs+MSTCN & 2.351 & 0.896 & 15.620 & 3.343  \\
U$^2$-LSTM+MSTCN  & 2.506 & 0.908 & 15.831 & 3.422  \\
Stage 1      & 2.581 & \textbf{0.912} & 16.094 & 3.453  \\
Stage 1+Stage 2      & \textbf{2.590} & \textbf{0.912} & \textbf{16.230} & \textbf{3.512}  \\ \hline
\end{tabular}
}
\end{table}

\subsection{Subjective and Objective Results on DNS-4 Challenge}
\label{ssec:performance}
In Table 2, we compare the proposed FB-MSTCN with the baseline model NSNet \cite{R20}, and present the evaluated results of subjective speech quality (SIG), background noise quality (BAK), overall audio quality (OVRL) with ITU-T P.835 criterion, and the objective Word Accuracy (WAcc). The results show that the FB-MSTCN model exhibits a very significant performance advantage and outperforms the baseline model overall by 0.59 OVRL MOS and 4.0$\%$ WAcc. In addition, it should be noted that our model is trained on multilingual data, while the blind test set of DNS-4 is only in English, so the FB-MSTCN model can achieve better performance if we train it for English only.

\begin{table}[th]
\caption{The averaged subjective and objective results of blind test set on DNS-4 challenge}
  \label{tab:The objective results for different models}
  \centering

\begin{tabular}{ccccc}
\hline
Method         & SIG  & BAK  & OVRL & WAcc \\ \hline
Noisy          & 4.29 & 2.15 & 2.63 & 0.72 \\
NSNet          & 3.62 & 3.93 & 3.26 & 0.63 \\
FB-MSTCN(Pro.) & 4.10 & 4.46 & 3.85 & 0.67 \\ \hline
\end{tabular}

\end{table}

To test whether the algorithm can run in real time, we evaluate the complexity and latency of the FB-MSTCN model. In this study, the proposed FB-MSTCN model has 29.9 M parameters and requires 12.5 G multiply-accumulate operations (MAC)  per second. The processing frame size is 20 ms with 10 ms overlap between frames. The average processing time per frame is 4.52 ms on the Intel i5-6400 CPU clocked at the fundamental frequency (2.7 GHz). The total algorithmic latency is 30 ms, which fully meets the requirement of real-time processing of DNS-4 challenge.




\section{Conclusion}
\label{sec:Conclusion}

For the full-band speech enhancement task in real-time scenarios, this paper proposes a new solution of extraction-interpolation. Through interval sampling, the difficult modeling problem of full-band spectrum can be effectively simplified into a modeling problem of three-channel wide-band spectrum. The proposed two-stage model, FB-MSTCN, further decomposes the enhancement problem of each wide-band spectrum  into a two-step optimization problem of `masking + compensation'. The experimental results show that the proposed method can not only meet the real-time requirements of DNS-4, but also has excellent performance in recovering the spectral details of full-band speech and suppressing the residual noise. On the blind test set of DNS-4 Track1, FB-MSTCN model has achieved very competitive performance, ranking third in the comprehensive evaluation results of MOS and WAcc.


\bibliographystyle{IEEEbib}
\bibliography{strings}

\end{document}